\input epsf
\documentstyle[12pt]{article}
        {\newpage
        \setcounter{page}{1}}

\renewcommand{\title}[1]{%
        {\begin{center}
        \Large\bf #1
        \end{center}}
        \vskip .3in}

\renewcommand{\author}[1]{%
        {\begin{center}
        #1
        \end{center}}}

\renewcommand{\abstract}[1]{%
        \begin{center}%
        {\vspace{1em}\vspace{0pt}\bf Abstract}%
        \end{center}%
        \noindent #1}

\renewcommand{\date}[1]{%
        \begin{center}%
        #1%
        \end{center}}

        {\end{thebibliography}}

\makeatletter
        \@addtoreset{equation}{section}%
\makeatother

\newcommand{\beq}{\begin{eqnarray}}
\newcommand{\eeq}{\end{eqnarray}}

\newcommand{\mybar}[1]%
        {\kern 0.8pt\overline{\kern -0.8pt#1\kern -0.8pt}\kern 0.8pt}
\newcommand{\sla}[1]%
        {\raise.15ex\hbox{$/$}\kern-.57em #1}
\newcommand{\roughly}[1]%
        {\mathrel{\raise.3ex\hbox{$#1$\kern-.75em\lower1ex\hbox{$\sim$}}}}

\newcommand{\jref}[4]{{\it #1} {\bf #2}, #3 (#4)}

\newcommand{\NPB}[3]{\jref{Nucl.\ Phys.}{B#1}{#2}{#3}}

\newcommand{\PRD}[3]{\jref{Phys.\ Rev.}{D#1}{#2}{#3}}

\newcommand{\PRL}[3]{\jref{Phys.\ Rev.\ Lett.}{#1}{#2}{#3}}

\begin{document}

\begin{titlepage}
\begin{center}
{\hbox to\hsize{ \hfill   MIT-CTP-2653}}
 
\bigskip
\bigskip
\bigskip
\vskip.2in

{\Large \bf  New  Mechanisms of Gauge-Mediated Supersymmetry 
Breaking  \footnote{Talk  presented at Rencontres de Moriond ``Electroweak
Interactions and Unified Theories'', Les Arcs France, March 15-22, 1997, on
work by L. Randall  \cite{mi} and C. Cs\'aki, L. Randall,
and W. Skiba \cite{mic}.}} \\

\bigskip
\bigskip
\bigskip
\vskip.2in
{\bf  Lisa Randall}\\

\vskip.2in

{ 
 Center for Theoretical Physics

Laboratory for Nuclear Science and Department of Physics

Massachusetts Institute of Technology

Cambridge,  MA 02139, USA

 }

\smallskip

{\tt  lisa@ctptop.mit.edu}

\vspace{1.5cm}
{\bf Abstract}\\
\end{center}

\bigskip

 We review the motivation for Gauge-Mediated Supersymmetry
Breaking and discuss some recent advances.

\bigskip

\end{titlepage}

\section{Introduction}
   What breaks
supersymmetry  is still an open question. In fact, it is 
  really two questions: what breaks supersymmetry and how is
it communciated to the visible world?
From the vantage point of what will be seen in experiments
this is an important question; the mass spectrum of superpartners
depends crucially on the mechanism of
communicating supersymmetry breaking.
 Perhaps the most compelling theory of supersymmetry
breaking would unify the two answers; thus would
be true in a scenario in which the visible world
and the supersymmetry breaking sector interact directly via
gauge and Yukawa couplings.
Some ideas along these lines have been developed recently
\cite{ratt,mur}, but in many models, the
sectors are  separated.
This is true both for supergravity and most messenger sector models. 

Supergravity mediation is still probably the scenario voted
most likely to succeed.  It has the virtue of simplicity (at
least when you don't ask too many questions).
 Soft supersymmetry breaking operators can be thought of
as derived from spurions:  {\bf  $S=\ldots+\theta \theta F$}
 \[
A_{ij}\left ({1 \over M_{Pl}}\right)^2\int d^4 \theta S^\dagger S Q_i^\dagger Q_j
\Rightarrow  m^{2}_{\tilde{Q}_{ij}}=A_{ij}\left({F_S \over M_{Pl}}\right)^2
\]
\[
{1 \over M_{Pl}} \int d^2 \theta S W_\alpha W^\alpha \Rightarrow m_{1/2}\sim  {F_S \over M_{Pl}}
\]
 Gravitational interactions guarantee that all    superpartners
are massive  and the relation
 $m_{\tilde{Q}} \sim m_{1/2}$ is automatic if there is a singlet
with a nonvanishing $F$ term which couples in the gauge kinetic term.
 If $A_{ij}\propto I$, there is  no {  dangerous} flavor violation.
However, it is difficult to understand why this should be the case. 
After all, we know the higher dimension operators in the Kahler
potential which contribute to the squark masses must be flavor
dependent since they need to absorb the divergence in the
flavor-dependent one-loop diagram which renormalizes the squark
mass.
 Unless counterterms   violate flavor  only through
terms proportional to Yukawas, which requires
some sort of underlying GIM mechanism \cite{hr}, one expects  flavor violation
in the  squark masses.
 This would permit dangerous flavor changing neutral
current effects, for which strong constraints exist.

 An advantageous alternative would have automatic
flavor conservation. The idea
 behind gauge-mediated models is that
superpartner masses are generated through gauge interactions.
 Scalar partners with the same quantum numbers have
the same mass; flavor symmetry is automatic. The relation
  $m_s\sim m_{1/2}$ is also automatic in the original models.
The mass relations for the scalars in these models
guarantees degeneracy among states with the same quantum
numbers as the masses are roughly
 $m_{\tilde{Q}}^2\sim g_3^2 \Lambda^2$,
 $m_{\tilde{L}}^2\sim g_2^2 \Lambda^2$,
 $m_{\tilde{e_R}}^2\sim g_1^2 \Lambda^2$,
 $m_{\tilde{g}}\sim g_3 \Lambda$, etc.
These are distinctive (testable) predictions
for the spectrum (although renormalization group
running can modify this spectrum).  Similarly,
the gauginos have a mass depending on their gauge charge.
In any particular model, one can generally find relations
among the scalar and fermionic superpartners, although these are more
model-dependent.

This seems a very nice and compelling picture.  The only
problem is that we need equally compelling models.  Since
mechanisms of dynamical supersymmetry breaking are increasingly
well understood in light of the many new tools which are currently
available for solving the infrared dynamics, it would
appear a simple task. However, we need to
 { communicate} supersymmetry breaking
to fields carrying only standard model  gauge charge.

One might be tempted to compare the relative virtues
of supergravity and gauge-mediation. 
    As in SUGRA, one can generate gaugino and sfermion masses
of about the same size.
In this case, the scalar partners
can be calculated perturbatively.
In the models in which a singlet is responsible for
communicating supersymmetry breaking to messenger quarks,
one finds that the
mass  squared for the scalars is positive. And of course,
the chief advantage is that 
because  the masses only depend on gauge charge,
scalars with the same quantum numbers are automatically degenerate
and the FCNC problem is automatically solved.

However, there are disadvantages to many of the existing models.
One problem is that the simplest models generally have a true
minimum in which QCD is broken \cite{broken}.
This is readily seen schematically by observing
that a superpotential with a nonvanishing $F_S$,
 $W=S F_S+S Q \bar{Q}$, would want  $Q \bar{Q}$ to  be nonzero.

Another problem which is of major importance for the credibility
of the models is that they are complicated; it is difficult to
give a nonzero $F$ term to singlets in many models;
existence proof models do exist though, which
is important as well \cite{DNNS}.  Even in models
with singlets one requires a barrier to the bad vacuum just discussed.
The final sticking point in these models  is the $\mu$ problem,
to which we have yet to find a satisfactory solution.

In what follows I will present some suggestions I made for improving
models. There are new ideas (some presented in this conference \cite{ratt})
which I will not discuss.  I will first
review ideas for trying to eliminate fundamental singlets
which proved problematic.  I will then go on to discuss how
these models can work well in conjunction with a ``Mediator'' scenario
I suggest. I will follow this with a discussion of a very simple
model, namely ``Intermediary '' models, in which singlets
link the DSB and visible sectors.

\section{Mediator Models}

A very nice idea was to try to eliminate the fundamental
singlets and instead to have composite singlets and messengers
which participate directly in  the dynamical symmetry breaking.
This idea was dubbed dynamical messenger sector (DMS) and
was investigated by Poppitz and Trivedi \cite{pt1,pt2} and 
Arkani-Hamed, March-Russell,and Murayama \cite{ahmrm}.
In fact, this idea is reminiscent of old attempts to link
the sectors; the problem with these models was
that there were many fields charged under the standard model
and consequently there were dangerous Landau poles.
In the more recent DMS models, the dangerous
Landau poles were circumvented because of the presence of two
mass scales.
  
However, for the particular models investigated in \cite{pt1,ahmrm},
the existence of two mass scales is also the root of the demise
of these models. The problem is that there exists
an
 intermediate mass regime in which the effective
$S Tr M^2$ is nonzero.
  In the effective theory, the relevant two-loop
diagrams will have a divergent contribution,
indicating a logarithmically enhanced contribution
which is cut off by the  heavy mass scale.
   This contribution to the mass squared dominates, and has
the wrong sign. 
These models are ruled out since they predict negative
squark and slepton mass squared.

However, these models might be useful in conjunction with
a Mediator Model.
The mediator models assume a
 weakly gauged global symmetry $G_m$ in the DSB sector
with a SUSY breaking gaugino mass $M_m$.
There are also mediator fields,
 $T(m,5)$, $\bar{T}(\bar{m},\bar{5})$, which
transform under both the messenger and standard model
gauge groups and therefore ``talk'' to both sectors.
It is assumed that these fields have a mass $M_T T \bar{T}$. Although
it seems contrary to the spirit of dynamical models to
introduce a mass term, we have in mind ulitmately an underlying
theory which induces this mass dynamically; we will discuss
such a model below. For now, it is useful first to analyze
the model assuming the existence of such a mass, and
then to investigate its potential origin.

With this simple additional assumption, namely the existence
of massive mediator fields, one finds the gauginos of the
visible sector are given a mass. No singlet coupled to
messenger quarks carrying standard model gauge charge
is required. This mass arises at three loops \cite{mi},
and involves exchange of $G_m$ gauginos, $T$ fermions, and
$T$ squarks. The necessary helicity flips are provided
by the gaugino and $T$ mass terms.  It is important
to note that so long as $M_m$, the messenger gaugino mass,
is less than $M_T$, the standard model gaugino mass is
independent of $M_T$, and is therefore not suppressed by
$M_T$ factors.

The scalar mass squared, on the other hand, arises at four loops.
Although it is too difficult to evaluate the four-loop
diagrams exactly, one can identify the dominant contribution.
This contribution can be understood as follows.  From the
point of view of the original DSB model, the $T$'s play
the role of squarks in the DMS models. Therefore, at ,
they receive a logarithmically enhanced {\it negative} contribution
to the squared scalar $T$ mass. This does not create tachyons
however, since by assumption the $T$ fields have a tree-level mass.
Now the $STr M^2$ over the $T$ fields is negative, which in
turn yields {\it positive} squark and slepton masses.

As can be deduced from the above, the squarks and sleptons
will be relatively heavy in these models. This makes
the models less natural, but is ultimately subject to experimental
verification.  The most generic signature for gauge mediated
models (that is, one which does not rely on a low supersymmetry
breaking scale and photon signatures) is the spectrum.
\[m_{i,1/2}={\alpha_i \over 4 \pi}{F \over S}
\]
For models based on a singlet coupling to messenger quarks, one finds
the ratio of scalar to gaugino mass is generally of order
unity, determined by the representation of the messengers, and is
\[
{m \over m_{i,1/2}}\approx \sqrt{2 k_i}
\]
This last relation is of course model-dependent; in the mediator models
 this relation is lost, and the scalars are relatively heavy.
 For mediator models  \footnote{We thank Asad Naqvi for these results.}:
 \[m_{i,1/2}={3 \over 4} N_f {\alpha_i \alpha_m^2 \over (4 \pi)^3} k_m
{F \over M} Log\left({M \over M_T}\right)^2
\]
where $M$ is the heavy messenger scale.
\[
{m_i \over m_{i,1/2}}\approx {50 \over N_f} \left({k_i \over k_m}\right)^{1/2}
\left[{\beta Log \left({M \over m_f}\right)^2-N_f \over Log\left({M\over M_T}\right)^2 } \right]^{1/2}{1 \over \alpha_m}>1
\]
where $\beta$ is of order unity.
Because these results are higher loop, they are more complicated.
The important feature to take away is the larger ratio of scalar
to gaugino mass.
If $N_f$, the number of flavors charged under the mediator group,
is relatively large as it tends to be in DMS models (in order
to have a sufficiently large global group which can be gauged),
this spectrum can be acceptable.

This potentially problematic feature of the spectrum
is not generic to all models which permit mass terms.
The mediator models communicated supersymmetry 
breaking through gauge and no superpotential interactions
(coupling to mediator fields). We next 
 a model without messenger gauge interactions.
There is a heavy singlet field which serves merely to induce
a higher dimensional operator which couples the DSM and messenger
sectors. 

\section{Intermediary Models}

The model works most simply with the following assumptions.
There is a vector multiplet, $V$ and $\bar{V}$, in the hidden sector.
There are singlet fields  $S$ and $\bar{S}$.
The superpotential includes the terms
\[ W =...+S V \bar{V}+M_S S \bar{S}+\bar{S} Q \bar{Q}+m_Q Q \bar{Q}
\]
One can integrate out the heaviest field, $S$, to generate
the following superpotential.
\[ W=...+{V \bar{V} Q \bar{Q} \over M_S} \]
 If $V \bar{V}$ has a nonvanishing $F$-term 
 the phenomenology  is like  usual gauge-mediated models.
 No complicated superpotential to generate a singlet $F$-term
is required.
This is a simple generic mechanism for communicating supersymmetry
breaking to messenger quarks and consequently to the visible sector.

\section{Composite Models}

We now return to the question of inducing the required
mass terms for the meditor and intermediary models
dynamically. This is considered in more detail in \cite{mic}.
We first present a composite mediator model.
Recall, all that is required
in this case is  $M_T \bar{T} T$, 
where the mediator group is assumed to be
SU(3) and under $SU(5) \times SU(3)$,
the fields transform as $T(5,3)$, $\bar{T}(\bar{5},\bar{3})$.
We assume the following microscopic theory.
\[
\begin{array}{c|cccc}
&SU(5)_{SM} &SU(3)_m &SP(4) & SP(4)' \\ \hline
T_5 & 5 & 1&4 &1 \\
T_3 & 1& 3& 4 &1 \\
\bar{T}_{\bar{5}} & \bar{5} & 1 & 1 & 4 \\
\bar{T}_{\bar{3}} & 1 & \bar{3} & 1 & 4 \\
\end{array}
\]


The bound states after the SP groups confine are
\[
\big (T_5^2\big)({10},1)\ \ \big(T_3\big)^2(1,\bar{3})  \ \ 
\big(\bar{T}_{\bar{5}}^2\big )(\bar{10},1)\ \ 
\big (T_{\bar{3}}^2\big )(1,3)
\]
\[ 
T\equiv \big(T_5 T_3\big) (5,3)\ \ \bar{T} \equiv\big(\bar{T_{\bar{5}}}
\bar{T_{\bar{3}}}\big)(\bar{5}, \bar{3})
\]
With the inclusion of appropriate tree-level terms in the superpotential,
the potential including the dynamical terms is
\[
W=Pf M +Pf M'+\lambda_1(T_5^2)(\bar{T_{\bar{5}}}^2)+
\lambda_2  T \bar{T}+
\lambda_3 (T_3^2)(\bar{T_{\bar{3}}}^2)
\]
which provides the necessary mass term.

One can also derive a composite intermediary model.
Here the subtlety is to derive an unsuppressed Yukawa coupling
in a composite model. Unless there are new dynamical terms
in the superpotential, tree-level terms to produce a Yukawa
vs. a mass term would have at least one higher dimension,
and therefore be suppressed by $M_P$.  We avoid
this by employing the dynamical Yukawa coupling which is
present in the confined theory.

 Recall that for intermediary models,
we need $M_S S \bar{S}+\lambda S Q \bar{Q}+M_Q Q \bar{Q}$.
Here we will assume the $V$ mass is unnecessary. In
Ref. \cite{mic}, it is shown how to extend this model
to include a $V$ mass.
The model we use is
\[
\begin{array}{c|ccc|c}
&SU(4)_{gauged}&SU(5) & SU(5) & SU(5)_D \\  \hline
q & 4 & 5 &1& 5 \\
\bar{q} & \bar{4} & 1 & \bar{5} & \bar{5}\\
\end{array}
\]
The confined spectrum for this model is
\[
M(5, \bar{5})\to\Sigma(24)+S(1)
\]
\[B(\bar{5},1) \to \bar{Q}(\bar{5})
\]
\[\bar{B}(1,5) \to \bar{Q}(5)
\]
and the superpotential is
\[
W=W_{tree}+det M - B M \bar{B}
\to W_{tree}
- Q \bar{Q} S - Q \Sigma \bar{Q}+det M 
\]
\[
W_{tree}=\bar{S} q \bar{q}+{1 \over M_{Pl}^5} q^4 \bar{q}^4+{1 \over M_{Pl}}
q \bar{q} q \bar{q}
\to \Lambda S \bar{S}+{\Lambda^6 \over M_{Pl}^5} Q \bar{Q}+{\Lambda^2 \over M_{Pl}}\Sigma^2
\]
There are  several constraints which must be imposed
on this model. These constraints
are discussed in greater detail in \cite{mic}.
We require that $M_Q>V F_V/M_S$ to avoid a tachyon, that the gravitino mass
is sufficiently low that gravity mediated effects
are suppressed and FCNC are avoided, consistency of the
perturbative expansion, and the correct scale for
the gaugino masses. An example of a viable solution
to these constraints is
\[
M_S \sim \Lambda\sim 10^{16} {\rm GeV} \ \ M_Q \sim {\Lambda^6 \over M_{Pl}^5}\sim 10^4 {\rm GeV}
\]
A high scale is probably necessary in any case in order to avoid
a dangerous Landau pole. These constraints are
discussed more extensively in Ref. \cite{mic}.

\section{Conclusions}

To conclude, the issue of supergravity vs. gauge mediation is
yet to be resolved and progress is still occurring.
It is conceivable that  FCNC constraints
will  motivate the correct choice,
as might simplicity.  The new developments in understanding
nonperturbative gauge dynamics should help unravel the
underlying structure.  In addition, the phenomenologically
distinctive signatures should play an important role.
It is worth noting that in these models, the ultimate
scale of supersymmetry breaking is generally sufficiently
high that one would not expect decays to gravitinos in the
detector. For both classes of models, it is the
distinctive spectrum that is the best signature.

\section{Acknowledgements}
I thank  Csaba Cs\'aki for comments on the manuscript
and Asad Naqvi for investigating some aspects of the phenomenology
of these models.
I also would like to acknowledge my collaborators
Csaba Cs\'aki  and Witold Skiba on the composite models.
 This work is  supported in part by DOE under cooperative
agreement \#DE-FC02-94ER40818, NSF Young Investigator
Award, Alfred P. Sloan Foundation Fellowship, DOE
Outstanding Investigator Award.



\begin{thebibliography}{99}

\bibitem{mi} L. Randall, hep-ph/9612426.

\bibitem{mic} C. Cs\'aki, L. Randall, and W. Skiba, MIT-CTP-2654
                                                                                    

\bibitem{DNNS} M. Dine, A.E. Nelson, and Y. Shirman, 
              \PRD{51}{1362}{1995}, hep-ph/9408384;
               M. Dine, A. Nelson, Y. Nir, and Y. Shirman,
               \PRD{53}{2658}{1996}, hep-ph/9507378.


 

\bibitem{broken} I. Dasgupta, B.A. Dobrescu, and L. Randall,
             \NPB{483}{95}{1997}, hep-ph/9507487;
             N. Arkani-Hamed, C. D. Carone, L. J. Hall, H. Murayama,
             \PRD{54}{7032}{1996}, hep-ph/9607298.

\bibitem{hr} L. Hall and L. Randall, \PRL{65}{2939}{1990}.

\bibitem{pt1} E. Poppitz and S. Trivedi, \PRD{55}{5508}{1997}, hep-ph/9609529

\bibitem{pt2} E. Poppitz and S. Trivedi,
             hep-ph/9703246.

\bibitem{ahmrm} N. Arkani-Hamed, J. March-Russell, and H. Murayama,
              hep-ph/9701286.


\bibitem{mur}  H. Murayama, hep-ph/9705271

\bibitem{ratt}
               S. Dimopoulos, G. Dvali, R. Rattazzi, and G. Giudice,
               hep-ph/9705307.





 
\end{thebibliography}
\end{document}